\documentclass[10pt,nohypertex,nofootinbib,onecolumn,prd,aps,superscriptaddress,notitlepage]{revtex4-2}

\usepackage[english]{babel}
\usepackage[T1]{fontenc}
\usepackage{mathrsfs}
\usepackage{amssymb, bm}
\usepackage{amsmath, amsthm}
\usepackage{hyperref}
\usepackage{enumerate}
\usepackage{color}
\usepackage{mathrsfs}
\usepackage{graphicx}
\usepackage{bm,natbib,url,textcase}
\usepackage{comment}
\numberwithin{equation}{section}
\usepackage{accents}
\usepackage{breakurl}

\allowdisplaybreaks

\hypersetup{
 colorlinks=true, % false: boxed links; true: colored links
 linkcolor=blue, % the colour of internal links
 citecolor=magenta, % the colour of links to the bibliography
 urlcolor=cyan % colour of external links
}

\definecolor{pinegreen}{rgb}{0.0, 0.47, 0.44}

\usepackage{tikz}

\definecolor{purple}{rgb}{1,0,1}
\definecolor{lime}{HTML}{a6CE39} % needs xcolor

\newcommand{\orcidicon}{%
	\begin{tikzpicture}
		\draw[lime, fill=lime] (0,0) 
		circle [radius=0.15] 
		node[white] {{\fontfamily{qag}\selectfont \tiny ID}};
		\draw[white, fill=white] (-0.0625,0.095) 
		circle [radius=0.007];
	\end{tikzpicture}	\hspace{-2mm}
}

\newcommand\orcidMarcello{{\href{https://orcid.org/0000-0003-0397-2705}{\orcidicon}}}

\newcommand\orcidDaniele{{\href{https://orcid.org/0000-0003-4379-2549}{\orcidicon}}}
\newcommand\orcidSalvatore{{\href{https://orcid.org/0000-0003-4886-2024}{\orcidicon}}}

\def\nn{\nonumber}

\begin{document}

\title{General analysis of Noether symmetries in Horndeski gravity}

\author{Marcello Miranda \orcidMarcello}
\email{marcello.miranda@unina.it}
\affiliation{Scuola Superiore Meridionale, Largo San Marcellino 10, 
I-80138, Napoli, Italy}
\affiliation{Istituto Nazionale di Fisica Nucleare, Sezione di Napoli, Complesso Universitario Monte Sant'Angelo, Edificio G, Via Cinthia, I-80126, Napoli, Italy}
%=================================================================
\author{Salvatore Capozziello \orcidSalvatore}
\email{capozziello@na.infn.it}
\affiliation{Scuola Superiore Meridionale, Largo San Marcellino 10, I-80138, Napoli, Italy}
\affiliation{Istituto Nazionale di Fisica Nucleare, Sezione di Napoli, Complesso Universitario Monte Sant'Angelo, Edificio G, Via Cinthia, I-80126, Napoli, Italy}
\affiliation{Dipartimento di Fisica ``E. Pancini'', Universit\`{a} di Napoli ``Federico II'', Complesso Universitario Monte Sant'Angelo, Edificio G, Via Cinthia, I-80126, Napoli, Italy}
%=================================================================
%=================================================================
\author{Daniele Vernieri \orcidDaniele}
\email{daniele.vernieri@unina.it}
\affiliation{Dipartimento di Fisica ``E. Pancini'', Universit\`{a} di Napoli ``Federico II'', Complesso Universitario Monte Sant'Angelo, Edificio G, Via Cinthia, I-80126, Napoli, Italy}
\affiliation{Istituto Nazionale di Fisica Nucleare, Sezione di Napoli, Complesso Universitario Monte Sant'Angelo, Edificio G, Via Cinthia, I-80126, Napoli, Italy}
\affiliation{Scuola Superiore Meridionale, Largo San Marcellino 10, I-80138, Napoli, Italy}
%=================================================================

\begin{abstract}
    We explore Noether symmetries of Horndeski gravity, extending the classification of general scalar-tensor theories. Starting from the minimally coupled scalar field and the first-generation scalar-tensor gravity, the discussion is generalised to kinetic gravity braiding and Horndeski gravity. We highlight the main findings by focusing on the non-minimally coupled Gauss-Bonnet term and the extended cuscuton model. Finally, we discuss how the presence of matter can influence Noether symmetries. It turns out that the selected Horndeski functions are unchanged with respect to the vacuum case.
\end{abstract}

\date{\today}

\maketitle

%
%
%

%
%
%

%%%%%%%%%%%%%%%%%%%%%%%%%%%%%%%%%%%%%%%%%%%%%%
\section{Introduction} \label{sec:intro}
\setcounter{equation}{0}
%%%%%%%%%%%%%%%%%%%%%%%%%%%%%%%%%%%%%%%%%%%%%%

The interest in understanding the fundamental nature of gravity has always been one of the main pursuits in Physics. Although General Relativity (GR) has passed several astrophysical and cosmological observational tests, and the $\Lambda$-Cold Dark Matter ($\Lambda$-CDM) model is considered the current standard model of cosmology, our understanding of gravity exhibits numerous significant limitations in both cosmology and quantum theory. The observed astrophysical and cosmological anomalies are addressed to the presence of the so-called \textit{dark fluids}, and they constitute the major part of the total energy-matter content of the Universe~\cite{Planck:2018vyg, Akrami:2018vks}. In particular, dark energy is responsible for the late-time cosmic accelerated expansion, while dark matter is, for instance, associated with the flatness of galaxy rotation curves. In the $\Lambda$-CDM model, dark energy is well described by a constant $\Lambda$ (\textit{i.e.}, the cosmological constant). However, this constant is manually plugged into GR equations without a solid theoretical explanation. Moreover, its observed magnitude is extremely small compared to the zero-point energy obtained from the quantum field theory. Therefore, the presence of several shortcomings prevents GR from being considered a complete theory, able to describe all the gravitational phenomena comprehensively. 

The current depicted scenario represents one of the reasons compelling the scientific community to explore novel approaches, based on the idea that GR should be modified or extended, to guarantee an exhaustive theory of gravitation~\cite{Joyce:2014kja}.
Based on this mindset, a huge plethora of alternative theories aim to overthrow GR from its status as \textit{The Theory of Gravity}~\cite{Capozziello:2011et, Clifton:2011jh, Heisenberg:2018vsk}. Furthermore, scientific advancements, within the development of new technologies for gravitational experiments, enable increasingly sensitive and stringent tests, probing gravity across different scales and energetic regimes, allowing us to discriminate among theories that would otherwise be observationally degenerate~\cite{LIGOScientific:2017ync, LIGOScientific:2017zic, NANOGrav:2023gor, DESI:2024uvr}.

One of the largest classes of extended theories of gravity is characterised by an additional scalar field. The crucial role of scalar fields in cosmology is widely recognised. Their versatility allows them to address various missing puzzles in our comprehension of the Universe's evolution at different eras: from early-time issues, attempting to solve the classical initial cosmic singularity and describing the inflation mechanism, to late-time cosmology, providing a dynamical description of the dark energy. The most famous and traditional formulations belong to the k-essence model, the scalar-tensor theories, as well as the Brans-Dicke theory, and the $f(R)$ gravity~\cite{Sotiriou:2008rp}, which admits a scalar-tensor formulation. 

Over time, the research of a more general theory has produced newer and more complex formulations, giving birth to theories like kinetic gravity braiding~\cite{Deffayet:2010qz, Gomes:2013ema}, Horndeski gravity~\cite{Horndeski:1974wa, Deffayet:2011gz, Kobayashi:2011nu} and beyond~\cite{Gleyzes:2014dya, Gleyzes:2014qga, BenAchour:2016cay, Langlois:2018dxi, Kobayashi:2019hrl}, yielding different novel contributions to the field equations.
In particular, the mentioned theories share the same characteristic: an additional propagating scalar degree of freedom associated with the scalar field, avoiding Ostrogradsky instability. This is precisely the criterion used in searching for more general and general theories with an additional scalar field. Nevertheless, more recent theories exhibit field equations with higher-order time derivatives, which are non-linearly depending on second-order time derivatives. Horndeski's theory is the most general scalar-tensor theory of gravity with second-order field equations avoiding Ostrogradsky instability.

It is well known that GR field equations depend linearly on second-order partial derivatives irrespective of the chosen background. This can be seen as a preferred theoretical framework for the equations of the dynamical field. However, if one wants to preserve this property, the \textit{final scalar-tensor theory} would be strongly constrained to a well-known subclass of Horndeski gravity, often called \textit{reduced} or \textit{viable} Horndeski. Precisely, the adjective \textit{viable} refers to the remarkable characteristic of the theory of having tensor perturbations propagating at the speed of light on dynamic backgrounds\footnote{This concerns the observational constraints placed on Horndeski gravity by the multi-messenger event GW170817/GRB170817A~\cite{LIGOScientific:2017ync, LIGOScientific:2017zic}.}, in a covariant way~\cite{Creminelli:2017sry, Baker:2017hug, Bettoni:2016mij, Andreou:2019ikc}. This physical constraint significantly increased the interest for these \textit{modern scalar-tensor theories}. Another aspect that makes these theories worthy of a deeper investigation is that viable Horndeski represents the \textit{smallest} class among all the generalised scalar-tensor theories containing models propagating only two tensor degrees of freedom\footnote{Up to (invertible) disformal transformations, or generalised disformal transformations, which do not increase the number of degrees of freedom of the theory~\cite{Bettoni:2013diz}.}. They are known in the literature as the cuscuton~\cite{Afshordi:2006ad, Afshordi:2007yx, Afshordi:2009tt} and, in general, the extended cuscuton model~\cite{Iyonaga:2018vnu, Iyonaga:2020bmm, Miranda:2022brj}. 

Additionally, moving through a different theoretical framework, it has been shown that viable Horndeski is the only theory carrying a \textit{general relativistic Newtonian} fluid interpretation\footnote{The used terminology refers to the traditional fluid dynamics and continuum mechanics. Linear constitutive relations characterise Newtonian fluids, while non-Newtonian fluids have more complex rheological behaviour. Newtonian fluid modelling provides a foundational comprehension of fluids behaviour, and they are used as references for comparing and understanding the behaviour of non-Newtonian fluids.}, while a more complex effective fluid is associated with more elaborate models~\cite{Faraoni:2018qdr, Miranda:2022wkz}. 
A recent formalism with intriguing applications to Horndeski gravity is known as the first-order thermodynamics of modified gravity, the so-called first-order thermodynamics of modified gravity~\cite{Faraoni:2021lfc, Giusti:2021sku, Faraoni:2021jri, Faraoni:2022gry, Giardino:2023ygc, Miranda:2024dhw}. Its goal is the construction of a unified framework for the landscape of gravity theories, including GR and its generalisations.

Another paramount aspect is the issue of the local well-posedness of the field equations for Horndeski gravity and beyond, which remains an open problem (beyond the scope of this paper). It has been shown that, in the vacuum, only viable Horndeski admits a generalised harmonic gauge condition for which the theory is strongly hyperbolic when linearised around a generic weak-field background\footnote{Only in the absence of the kinetic gravity braiding term, the generalised harmonic gauge condition for the linearised theory arise by linearising a generalised harmonic gauge condition for the non-linear theory. There are ways to avoid this issue (\textit{i.e.}, considering spherical symmetric spacetime), but we want to address this argument to emphasise the actual importance of studying \textit{simpler} theories and analysing them in different frameworks and points of view. Lastly, it is worth highlighting that only in the vacuum this class can be seen as equivalent to a general k-essence model (since the non-minimal coupling function can be removed only by performing a conformal transformation and not by a scalar field redefinition). However, the equivalence between the Jordan and Einstein frame remains open to debate.}~\cite{Papallo:2017qvl, Kovacs:2020ywu}. 

For all the outlined reasons, the importance of deeply studying such a theory is evident, as it represents a critical step in the mathematical and physical understanding of more elaborate formulations. 

The presence of so many free unknown functions appearing in the Horndeski action makes it challenging to grasp the physical meaning of the theory, as well as to establish a classification for all the possible subclasses. Moreover, these free functions are usually set \textit{ad hoc} to face different problems, making it even more challenging to orient oneself among the infinite possibilities.

The goal of this paper is to classify the Horndeski models according to the Noether Symmetry Approach~\cite{Capozziello:1996bi, Bajardi:2022ypn}, extending previous works of the literature. On one side, the existence of symmetries allows us to solve the dynamics exactly; on the other, the Noether charge can always be related to some observable quantity. Unlike precedent works~\cite{Dimakis:2017zdu, Capozziello:2018gms}, a Lagrange multiplier will be used to keep the braiding function general and, at the same time, deal with point-like Lagrangian depending at most on the time first derivative of the configuration space\footnote{The use of Lagrange multiplier in finding Noether symmetries is present in different modified theories of gravity, for instance, $f(R)$ gravity~\cite{Capozziello:2008ch}, $f(\mathcal{G})$ gravity~\cite{Bajardi:2020osh, Bajardi:2020xfj}, and non-local gravity~\cite{Bajardi:2020mdp}.}. 

Since the presence (or absence) of Horndeski functions influences the result of the Noether Symmetry Approach in selecting the Lagrangian functional form, a brief review of all the scalar-tensor subclasses included in Horndeski gravity is provided. It allows us to appreciate the hierarchical structure of the scalar-tensor theories and their relation to the infinitesimal generators of the symmetries. For consistency, the modern Horndeski gravity \textit{nomenclature} is used throughout the entire work: $G_2$ for the k-essence contribution, $G_3$ for the kinetic braiding term, $G_4$ and $G_5$ for the non-minimally coupling functions of the Ricci scalar and the Einstein tensor, respectively.

It is worth highlighting that this approach reverses the usual application of the Noether theorem (see also Refs.~\cite{Vakili:2008ea, Capozziello:2012iea, Paliathanasis:2014rja}). The Noether Symmetry Approach constitutes a criterion to select the functional forms of the arbitrary Horndeski functions, by assuming the invariance under Noether point symmetries. Usually, the Noether theorem is used to obtain integrals of motion corresponding to transformations leaving the action invariant. An exhaustive and general discussion is presented in Ref.~\cite{Bajardi:2022ypn}. This method provided several exact solutions of the gravitational field equations, describing the time evolution of a spatially flat Friedman-Lema\^{i}tre-Robertson-Walker (FLRW) universe for the scalar-tensor and Gauss-Bonnet theory.
\newpage
Throughout this work, the following conventions are adopted: $g_{\mu\nu}= {\rm diag}(-1, a^2, a^2, a^2)$, being $a=a(t)$ the scale factor, the \textit{over-dot} represents the (total) time derivative, the scalar field and the kinetic term are denoted by $\phi=\phi(t)$ and $X=\tfrac{1}{2}\dot{\phi}^2$, respectively, and $8\pi G = c = \hbar = 1$ (reduced Planck units).\\

In Sec.~\ref{sec:kess}, the Noether Symmetry Approach is summarised and applied to the k-essence model. In Sec.~\ref{sec:nonmini}, we review the Noether classification of scalar-tensor theories with the \textit{traditional} non-minimally coupled scalar field to the Ricci scalar. In Sec.~\ref{sec:braiding}, we present the Noether Symmetry Approach specifically for kinetic braiding gravity (constituting our first achievement in generalising the Noether analysis) and generalised in Sec.~\ref{sec:horn} to Horndeski gravity. In Sec.~\ref{sec:models}, particular modified theories of gravity, Gauss-Bonnet gravity and the extended cuscuton model, are analysed in this framework. The form of Lagrangians is chosen to guarantee the existence of a Noether symmetry. Next, in Sec.~\ref{sec:matter}, we discuss the classification of Noether symmetries in the presence of matter. The final discussion is presented in Sec.~\ref{sec:disc}. 

\section{k-essence model}\label{sec:kess}
\setcounter{equation}{0}

Let us start this section by considering the case of a canonical kinetic term,
\begin{equation}\label{eq:canon_act}
    S=\int{d^4x\sqrt{-g}\left(R-\frac{1}{2}\nabla_{a}\phi\nabla^{a}\phi-V(\phi)\right)}\,.
\end{equation}
Evaluating the above action for the spatially flat FLRW universe, for which the Ricci scalar is $R=6\left(\tfrac{\dot{a}^2}{a^2}+\tfrac{\ddot{a}}{a}\right)$, the point-like Lagrangian turns out to be
\begin{align}\label{eq:canon_lagr}
    \mathcal{L}&= 6a\dot{a}^2+6a^2\ddot{a}+\frac{1}{2}a^3\dot{\phi}^2-a^3V\nn\\
    &=-6a\dot{a}^2+\frac{1}{2}a^3\dot{\phi}^2-a^3V\,,
\end{align}
by performing an integration by parts to eliminate the second derivative of the scale factor\footnote{The integration by parts is not necessary, in general. This step is equivalent to considering a \textit{second prolongation} of the infinitesimal generator of the Noether symmetry. However, while here the integration is sufficient to rewrite the point-like Lagrangian in a first-order canonical form (\textit{i.e.}, depending only on the first derivatives), it is not the case of kinetic gravity braiding and Horndeski gravity.}.

The infinitesimal generator of the Noether symmetry is written as follows
\begin{equation}\label{eq:inf_gen}
    \chi=\xi(t,a,\phi)\partial_{t}+\eta_{a}(t,a,\phi)\partial_{a}+\eta_{\phi}(t,a,\phi)\partial_{\phi}\,,
\end{equation}
then the \textit{first prolongation} corresponds to
\begin{equation}\label{eq:first_prolong}
    \chi^{[1]}=\chi+\big(\dot{\eta}_{a}-\dot{a}\,\dot{\xi}\,\big)\partial_{\dot{a}}+\big(\dot{\eta}_{\phi}-\dot{\phi}\,\dot{\xi}\,\big)\partial_{\dot{\phi}}\,.
\end{equation}
The existence of a Noether symmetry for the point-like Lagrangian~\eqref{eq:canon_lagr} is ensured by the following identity:
\begin{equation}\label{eq:noether}
    \chi^{[1]}\mathcal{L}+\dot{\xi}\mathcal{L}=\dot{\zeta}(t,a,\phi)\,,
\end{equation}
where $\dot{\zeta}$ is a generic function corresponding to the gauge freedom of the symmetry, and it can be safely set to zero (\textit{i.e.}, $\zeta=\mbox{cost}$), without losing generality. 
Then, using the Hamiltonian constraint corresponding to the first Friedmann equation,
\begin{equation}\label{eq:energy}
    \mathcal{L}-\frac{\partial\mathcal{L}}{\partial\dot{a}}\dot{a}-\frac{\partial\mathcal{L}}{\partial\dot{\phi}}\dot{\phi}=0\,,
\end{equation}
the associated conserved quantity reads as follows,
\begin{equation}\label{eq:conserved}
    \mathcal{J}=\zeta-\eta_{a}\frac{\partial\mathcal{L}}{\partial \dot{a}}-\eta_{\phi}\frac{\partial\mathcal{L}}{\partial \dot{\phi}}\,.
\end{equation}

Thus, the identity~Eq.\eqref{eq:noether} becomes a set of equations for the functions $\xi$, $\eta_a$, $\eta_\phi$, and $V$, by setting to zero the coefficients obtained by factorising all the time-derivative terms, \textit{i.e.}, $\dot{\phi}\,\dot{a}$, $\dot{\phi}^2\,\dot{a}$, $\dot{\phi}\,\dot{a}^2$, $\dot{\phi}^i$, $\dot{a}^i$, with $i={1,2,3}$. This yields the following configurations:
\begin{equation}\label{eq:canon_gen}
    \xi(t)= \xi_1 t + \xi_2\,,\quad \eta_a(a,\phi)=\frac{a}{3}\,\xi_1+\frac{c_a(\phi)}{\sqrt{a}}\,,\quad \eta_\phi(a,\phi)=\xi_0-8\frac{c_{a}(\phi)}{a^{3/2}}\,,
\end{equation}
with
\begin{align}
    {\rm\left(\,I\,\right)}:\quad&\begin{cases}
    \vspace{-10pt}\\
        c_a(\phi)=0\vspace{10pt}\\
        \xi_0\neq0\neq \xi_1\vspace{10pt}\\
        V(\phi)=V_0\exp{\left(-\frac{2\xi_1}{\xi_0}\phi\right)}\vspace{3pt}
    \end{cases}\qquad\qquad\quad
    {\rm\left(\,II\,\right)}:\quad\begin{cases}
    \vspace{-10pt}\\
        c_a(\phi)=c_1\exp{\left(\frac{\sqrt{3}}{4}\phi\right)}+c_2\exp{\left(-\frac{\sqrt{3}}{4}\phi\right)}\vspace{10pt}\\
        V(\phi)=V_0\exp{\left(-\frac{2\xi_1}{\xi_0}\phi\right)}\vspace{10pt}\\
        \begin{cases}
        \vspace{-10pt}\\
            c_1=0\neq\xi_1\vspace{10pt}\\
            \xi_0=\frac{4}{\sqrt{3}}\xi_1\vspace{3pt}
        \end{cases}
        \quad{\vee}\qquad
        \begin{cases}
        \vspace{-10pt}\\
            c_2=0\neq\xi_1\vspace{10pt}\\
            \xi_0=-\frac{4}{\sqrt{3}}\xi_1\vspace{3pt}
        \end{cases}\vspace{4pt}
    \end{cases}
\end{align}
or, in the case of internal symmetries\footnote{Notice that, in general, internal symmetries correspond to $\xi(t)=0$. However, at the level of the Noether identity~\eqref{eq:noether}, there is no difference between a vanishing and a constant $\xi$. This is because the Lagrangian does not explicitly depend on the time coordinate.
Then, $\chi$ does not have the $\xi$ component, and $\chi^{[1]}$ depends only on $\dot{\xi}$. Therefore, both $\xi=0$ and $\dot{\xi}=0$ correspond to internal symmetries.
Moreover, in cosmology, due to the energy constraint~\eqref{eq:energy}, $\xi$ is not present in the expression of the conserved scalar current associated with the Noether symmetry~\eqref{eq:conserved}.},
\begin{equation}
    {\rm\left(\,III\,\right)}:\quad\begin{cases}
    \vspace{-10pt}\\
        c_a(\phi)=c_1\exp{\left(\frac{\sqrt{3}}{4}\phi\right)}+c_2\exp{\left(-\frac{\sqrt{3}}{4}\phi\right)}\vspace{10pt}\\
        V(\phi)=V_0 \exp{\left(-\frac{\sqrt{3}}{2}\phi \right)} \left[c_2-c_1 \exp{\left(\frac{\sqrt{3}}{2}\phi \right)}\right]^2\vspace{10pt}\\
        \xi_0=0=\xi_1\vspace{4pt}
    \end{cases}
\end{equation}
where, $\xi_{0,1,2}$, $c_{1,2}$, and $V_0$ are arbitrary constants.

It is possible to verify that a phantom scalar field (\textit{i.e.}, in the case of a negative sign in front of the kinetic term) admits a Noether point symmetry only in correspondence with the first set of parameters.

In the case of a general (unknown) kinetic dependence (like the k-essence model),
\begin{equation}\label{eq:k-act}
    S=\int{d^4x\sqrt{-g}\left[R+G_{2}(\phi,X)\right]}\,,
\end{equation}
where, using the modern Horndeski gravity notation, $G_{2}$ is a generic function of the scalar field and the kinetic term, $X=-\frac{1}{2}\nabla_{a}\phi\nabla^{a}\phi$. In this case, the kinetic term can be treated as a new additional variable, and its definition must be included in the theory by using a Lagrange multiplier.
Adding a Lagrange multiplier does not change the field equations, and it is perfectly equivalent to considering the definition of $X$ from the beginning.

To apply the Noether Symmetry Approach it is necessary to write down the point-like Lagrangian of the theory in a canonical way and to split the identity guaranteeing the existence of the symmetry~\eqref{eq:noether} in a set of equations, by collecting all the time-derivative terms. The using of the Lagrange multiplier makes the process easier. The discussion is analogous to the one made for other theories of gravity~\cite{Capozziello:2008ch, Bajardi:2020xfj, Bajardi:2020osh}.
Thus, the point-like Lagrangian reads
\begin{equation}
    \mathcal{L}= -6a\dot{a}^2+a^3 G_{2}(\phi,X)+a^3\lambda\left(X-\frac{1}{2}\dot{\phi}^2\right).
\end{equation}
The variation with respect to $X$ gives $\lambda=-\partial_{X}G_{2}(\phi,X)$. Therefore, the previous Lagrangian can be rewritten as follows
\begin{equation}\label{eq:k-lagr}
    \mathcal{L}= -6a\dot{a}^2+a^3 G_{2}(\phi,X)-a^3\partial_{X}G_{2}(\phi,X)\left(X-\frac{1}{2}\dot{\phi}^2\right).
\end{equation}
The infinitesimal generator of the Noether symmetry is provided by
\begin{equation}\label{eq:chi}
    \chi=\xi(t,a,\phi,X)\partial_{t}+\eta_{a}(t,a,\phi,X)\partial_{a}+\eta_{\phi}(t,a,\phi,X)\partial_{\phi}+\eta_{X}(t,a,\phi,X)\partial_{X}\,,
\end{equation}
and, the \textit{first prolongation} corresponds to
\begin{equation}\label{eq:k-first-prolong}
    \chi^{[1]}=\chi+\big(\dot{\eta}_{a}-\dot{a}\,\dot{\xi}\,\big)\partial_{\dot{a}}+\big(\dot{\eta}_{\phi}-\dot{\phi}\,\dot{\xi}\,\big)\partial_{\dot{\phi}}+\big(\dot{\eta}_{X}-\dot{X}\,\dot{\xi}\,\big)\partial_{\dot{X}}\,.
\end{equation}

Following the same procedure described above, setting to zero the coefficients obtained by factorising all the time-derivative terms, the Noether Symmetry Approach yields the following configurations:
\begin{equation}\label{eq:k-gen}
    \xi(t)= \xi_1 t + \xi_2\,,\quad \eta_a(a)=\frac{a}{3}\,\xi_1\,,\quad \eta_\phi=\eta_\phi(\phi)\,,\quad \eta_X(\phi,X)=2X(\partial_{\phi}\eta_{\phi}-\xi_1)\,,
\end{equation}

\begin{align}\label{eq:k-eta}
    {\rm\left(\,I\,\right)}:\quad&\begin{cases}
        \vspace{-10pt}\\
        \xi_1\neq0\vspace{10pt}\\
        \eta_{\phi}(\phi)=\xi_1 \phi+\xi_0\vspace{10pt}\\
        G_{2}(\phi,X)=\dfrac{g_2(X)}{\xi_1\phi+\xi_0}\vspace{3pt}
    \end{cases}\\[10pt]
    {\rm\left(\,II\,\right)}:\quad&\begin{cases}
        \vspace{-10pt}\\
        \xi_1\neq0\vspace{10pt}\\
        \eta_{\phi}(\phi)\neq\xi_1 \phi+\xi_0\vspace{10pt}\\
        G_{2}(\phi,X)=\exp {\displaystyle\left(-\int _1^{\phi }\frac{2 \xi_1}{\eta_{\phi}(\varphi)}d\varphi\right)} g_2\left(X \exp {\displaystyle\left(\int _1^{\phi }\frac{2 \left(\xi_1-\partial_{\varphi}\eta_{\phi}(\varphi)\right)}{\eta_{\phi}(\varphi)}d\varphi\right)}\right)\vspace{3pt}
    \end{cases}
\end{align}
or, in the case of internal symmetries,
\begin{equation}
    {\rm\left(\,III\,\right)}:\quad\begin{cases}
        \vspace{-10pt}\\
        \xi_1=0\vspace{10pt}\\
        G_{2}(\phi,X)= g_2\left(\dfrac{X}{\eta_{\phi}^2(\phi)}\right)\vspace{3pt}
    \end{cases}
\end{equation}
where, $\xi_{0,1,2}$ are constants, and $g_2$ is an arbitrary function of $X$ times a factorised scalar field dependence (eventually constant). Notice that $g_{2}$ can be a linear function of its variable, \textit{a posteriori}. However, compared to the analysis done on the linear dependence from the beginning~\eqref{eq:canon_lagr}, the above results are over-constraining the theory.

The above characterisations can be rewritten so that the kinetic dependence is fully factorised from the pure scalar field one, $G_2(\phi,X)=h(\phi)\,g_2(X)$, by redefining the scalar field. Then, one obtain $\eta_{\phi}(\phi)=\xi_1 \phi+\xi_0$ or $\eta_{\phi}(\phi)=\xi_0$. However, imposing a factorised form of $G_2$ with a general $\phi$-component of the infinitesimal generator, one obtain $g_2(X)=c_0\,X^{c_\phi}$, $h(\phi)=\left(\xi_1-\partial_{\phi}\eta_{\phi}\right)^{-1}$, where $\eta_{\phi}$ is implicitly defined by a differential equation, $2\left(\xi1-\partial_{\phi}\eta_{\phi}\right)\left[(c_\phi-1)\xi_1-c_\phi\partial_\phi\eta_\phi\right]
-\eta_\phi\partial^2_\phi\eta_\phi=0$, with $c_\phi\neq1$ being a constant. For instance, the last equation is satisfied for $\eta_{\phi}=\left(\tfrac{c_{\phi}-1}{c_{\phi}}\right)\xi_1\phi+\xi_0$.

\section{Traditional non-minimally coupled scalar field}\label{sec:nonmini}
\setcounter{equation}{0}

The action of the first-generation scalar-tensor theory is
\begin{equation}
    S=\int{d^4x\sqrt{-g}\left[G_{4}(\phi)R-\frac{\omega(\phi)}{2}\nabla_{a}\phi\nabla^{a}\phi-V(\phi)\right]}\,,
\end{equation}
where, the non-minimal coupling $G_4$ is an arbitrary function of $\phi$.
The corresponding to the following point-like action is
\begin{align}\label{eq:scalartensor-lagr}
    \mathcal{L}=&\,6a\dot{a}^2G_{4}(\phi)+6a^2\ddot{a}\,G_{4}(\phi)+a^3 \frac{\omega(\phi)}{2}\dot{\phi}^2-a^3V(\phi)\nn\\
    =&-6a\dot{a}^2G_{4}(\phi)-6a^2\dot{a}\dot{\phi}\,\partial_{\phi}G_{4}(\phi)+a^3 \frac{\omega(\phi)}{2}\dot{\phi}^2-a^3V(\phi).
\end{align}

As done in the previous section, the configurations obtained turn out to be:
\begin{equation}\label{eq:scalartensor-gen}
    \xi(t)=\xi_1 t+\xi_2\,,\quad \eta_{a}(a,\phi)=-\frac{a}{3}\left(\xi_1+\frac{\partial_{\phi}V(\phi)}{V(\phi)}\eta_{\phi}(a,\phi)\right)\,,\quad\eta_{\phi}=\eta_{\phi}(a,\phi)
\end{equation}
\begin{align}
    {\rm\left(\,I\,\right)}:\quad&\begin{cases}
        \vspace{-10pt}\\
        V(\phi)=V_0 \, G_{4}(\phi)\vspace{10pt}\\
        \omega(\phi)=\omega_0\dfrac{\left(\partial_\phi G_4(\phi)\right)^2}{G_4(\phi)}\vspace{10pt}\\
        \eta_{\phi}(a,\phi)=\dfrac{{G_4}(\phi )}{\partial_\phi G_{4}(\phi )} \left[3 \xi_1\Big(2 \ln (a)+(\omega_0+4) \ln ({G_4}(\phi ))\Big)+\xi_0\right]\vspace{10pt}\\
        \xi_1=0\quad\vee\quad\omega_0=-\dfrac{8}{3}\quad\vee\quad\omega_0=-3\vspace{5pt}
    \end{cases}\\[10pt]
    {\rm\left(\,II\,\right)}:\quad&\begin{cases}
        {\,}\vspace{-9pt}\\
        V(\phi)=V_0 \, G_{4}(\phi)^{\,3/2}\vspace{10pt}\\
        \omega(\phi)=\omega_0\dfrac{\left(\partial_\phi G_4(\phi)\right)^2}{G_4(\phi)}\vspace{10pt}\\
        \eta_\phi (\phi)= -4\, \xi_1 \dfrac{G_4(\phi)}{\partial_\phi G_4(\phi)}\vspace{10pt}\\
        \xi_1\neq0\vspace{3pt}
    \end{cases}\\[10pt]
    {\rm\left(\,III\,\right)}:\quad&\begin{cases}
        \vspace{-5pt}\\
        V(\phi)=V_0 \, G_{4}(\phi)^{\,\gamma}\,,\quad \gamma\neq1\,,\tfrac{3}{2}\vspace{10pt}\\
         \omega(\phi)=\omega_{0}  \dfrac{\left(\partial_\phi G_4(\phi)\right)^2}{G_4(\phi)}\vspace{10pt}\\
         \eta_{\phi}(a,\phi)=\dfrac{2 \xi_1 }{(1-\gamma) }\dfrac{G_{4}(\phi)}{\partial_{\phi}G_{4}(\phi)}+\xi_{0}\, \dfrac{G_4(\phi )^{\alpha}}{ \partial_{\phi}G_{4}(\phi)} \,a^{\beta}\vspace{10pt}\\
         \alpha=\dfrac{2 \gamma ^2+3 (\gamma -1) \omega_{0}}{2 (3-2 \gamma )^2}\,,\quad\beta=\dfrac{3(1-\gamma)}{3-2\gamma} \gamma \vspace{3pt}\vspace{10pt}\\
        \xi_0=0\quad\vee\quad\omega_0=-3\quad\vee\quad\omega_0=\dfrac{4}{3} \,\gamma\,(\gamma -3) \vspace{5pt}
    \end{cases}
\end{align}
where, $\xi_{0,1,2}$, $\alpha$, $\beta$, $\gamma$, $\omega_0$, and $V_0$ are arbitrary constants.

Let us now consider a more general scalar-tensor theory,
\begin{equation}
    S=\int{d^4x\sqrt{-g}\left[G_{4}(\phi)R+G_{2}(\phi,X)\right]}\,.
\end{equation}
where the non-minimal coupling function $G_{4}$ is assumed to be not constant. 

As done in the previous section, excluding the linear case, the kinetic term can be treated as an additional variable, by including a Lagrange multiplier. After solving the Lagrange multiplier, $\lambda=-\partial_{X}G_{2}(\phi,X)$, the point-like Lagrangian reads as
\begin{equation}\label{eq:genscalartensor-lagr}
    \mathcal{L}= -6a\dot{a}^2G_{4}(\phi)-6a^2\dot{a}\dot{\phi}\,\partial_{\phi}G_{4}(\phi)+a^3 G_{2}(\phi,X)-a^3\partial_{X}G_{2}(\phi,X)\left(X-\frac{1}{2}\dot{\phi}^2\right),
\end{equation}
corresponding to the following classification,
\begin{equation}\label{eq:genscalartensor-gen}
    \xi(t)=\xi_1\,t+\xi_2\,,\quad\eta_{a}(a)=\frac{a}{3}(\xi_1-\phi_0)\,,\quad\eta_{\phi}(\phi)=\phi_0\frac{G_4(\phi)}{\partial_{\phi}G_{4}(\phi)}\,,\quad\eta_{X}(\phi,X)=2X\left(\partial_{\phi}\eta_{\phi}-\xi_1\right)
\end{equation}
with
\begin{align}
{\rm\left(\,I\,\right)}:\quad&\begin{cases}\label{eq:scalartensor-eta}
        \vspace{-10pt}\\
        \xi_1\neq0\vspace{10pt}\\
        \eta_{\phi}(\phi)=\xi_1 \phi+\xi_0\vspace{10pt}\\
        G_{4}(\phi)=c_{\phi}\,(\xi_1 \phi+\xi_0)^{\phi_0/\xi_1}\vspace{10pt}\\
        G_{2}(\phi,X)= g_2(X)(\xi_1 \phi+\xi_0)^{\frac{\phi_0}{\xi_1} -2}\vspace{3pt}\\
    \end{cases}\\[10pt]
    {\rm\left(\,II\,\right)}:\quad&\begin{cases}
    \vspace{-10pt}\\
        \eta_{\phi}(\phi)\neq\xi_1\phi+\xi_0\vspace{10pt}\\
        G_{2}(\phi,X)=G_4(\phi )^{1-\frac{2 \xi_1}{\phi_0}} g_2\left(X \dfrac{\partial_{\phi}G_4(\phi )^2}{ G_4(\phi )^{2-\frac{2 \xi_{1}}{\phi_0}}}\right)\vspace{3pt}\\
    \end{cases}
\end{align}
where, $\xi_{0,1,2}$, $c_{\phi}$, and $\phi_0$ are constants, and $g_2$ is an arbitrary function of $X$ times a factorised scalar field dependence (eventually constant). Notice that $g_{2}$ can be, in general, a linear function of its variable but the above results are over-constraining the theory compared to the analysis done on the linear dependence from the beginning.

The latter system can also be rewritten in terms of $\eta_{\phi}$, in the following way
\begin{equation}
    {\rm\left(\,II\,\right)}:\quad\begin{cases}
        \vspace{-7pt}\\
        \eta_{\phi}(\phi)\neq\xi_1 \phi+\xi_0\vspace{10pt}\\
        G_{4}(\phi)=c_{0}\,\int _1^{\phi }\exp\left({\int _1^{\varphi_2}\frac{\phi_0-\partial_{\varphi_1}\eta_{\phi}(\varphi_1)}{\eta_{\phi} (\varphi_1)}d\varphi_1}\right) d\varphi_2\vspace{10pt}\\
        G_{2}(\phi,X)=\exp \left(\int _1^{\phi }-\frac{2 \xi_1-\phi_0}{\eta_{\phi} (\varphi)}d\varphi\right) g_2\left(X \exp \left(\int _1^{\phi }\frac{2 \left(\xi_1-\partial_{\varphi}\eta_{\phi} (\varphi)\right)}{\eta_{\phi} (\varphi)}d\varphi\right)\right)\vspace{5pt}
    \end{cases}
\end{equation}

Unlike the result obtained for the k-essence model, $\eta_{\phi}$ is fully determined by the non-minimal coupling function. Moreover, the presence of the non-constant $G_4$ implies the absence of the pure shift-symmetry with respect to the scalar field. 
It is replaced by a \textit{generalised shift-symmetry} corresponding to $\eta_{\phi}=\xi_0$, $\xi(t)=\xi_1 \,t+\xi_2$, $\eta_{a}=a(\xi_1-\phi_0)/3$, and $G_{4}(\phi)=\exp\left(\tfrac{\phi_0}{\xi_0}\phi\right)$. Allowing $G_{4}$ to be constant, the shit-symmetry is obtained in correspondence with $\phi_0=0$ and $\xi_1=0$. Notice that, using the parameterisation~\eqref{eq:scalartensor-eta}, the case of the minimally-coupled scalar field~\eqref{eq:k-eta} is obtained by setting $\phi_0=0$. Indeed, as it will be clear in the next sections, $\phi_0$ is always associated with the $\phi$ dependence of the non-minimal coupling function $G_4$.

When, for the former case, the factorisation of $\phi$ and $X$ dependence is imposed, $G_2(\phi,X)=h(\phi)g_2(X)$, irrespectively of $\eta_{\phi}$ (\textit{i.e.}, assuming $\eta_{\phi}\neq\xi_1 \phi + \xi_0$ and $\eta_{\phi}\neq\xi_0$), it yields 
\begin{equation}
    G_2(\phi,X)=c_0 \left(\partial_{\phi}G_4(\phi )\right)^{2 c_\phi}\, G_4(\phi )^{\frac{2 (c_\phi-1) \xi_1}{\phi_0}+1-2 c_\phi}   \,X^{c_{\phi}}\,,
\end{equation}
where $c_\phi\neq1$, since we are excluding the linear case.

\section{Kinetic gravity braiding}\label{sec:braiding}
\setcounter{equation}{0}

Let us turn on the braiding term $G_3(\phi,X)$ inside the action, proportional to $\Box\phi=\nabla_{a}\nabla^{a}\phi$,
\begin{equation}\label{sec:braiding_act}
    S=\int{d^4x\sqrt{-g}\left[R+G_{2}(\phi,X)-G_{3}(\phi,X)\Box \phi\right]}\,.
\end{equation}

The nomenclature \textit{kinetic braiding} refers to the fact that the function $G_3$ must depend on the kinetic term to contribute in a non-trivial way compared to the k-essence part. Therefore, we are focusing on the case of $\partial_{X}G_{3}\neq0$. Indeed, if $G_3=G_3(\phi)$ the action~\eqref{sec:braiding_act} can be recast into the k-essence model by integrating by parts: $G_3(\phi)\Box\phi=-2X\partial_{\phi}G_3(\phi)\subseteq G_{2}(\phi,X)$, up to a total divergence.

In the past, this term yielded only a partial classification according to the Noether symmetries by making some assumptions on $X$ dependence of $G_3$~\cite{Capozziello:2018gms}. This lies in the presence of the D'Alambertian of the scalar field. Indeed, substituting $\Box\phi=-\left(\ddot{\phi}+3\tfrac{\dot{a}}{a}\dot{\phi}\right)$, the point-like Lagrangian associated with the braiding term is
\begin{equation}
    \mathcal{L}_{3}=a^3\,G_{3}(\phi,X)\left(\ddot{\phi}+3\frac{\dot{a}}{a}\dot{\phi}\right)\,.
\end{equation}
The term $a^3\,G_{3}(\phi,X)\ddot{\phi}$ cannot be transformed in a first-order Lagrangian by integrating by parts in general. However, the Noether symmetries analysis of this theory can be performed by taking into account the \textit{second prolongation} of the infinitesimal generator, $\chi^{[2]}=\chi^{[1]}+(\ddot{\eta}_\phi-\dot{\phi}\,\ddot{\xi}-2\ddot{\phi}\,\dot{\xi})\partial_{\ddot{\phi}}$, and implementing the Noether identity~\eqref{eq:noether}. The only way to recast the Lagrangian in a canonical form is by considering $X$ as a new independent variable of our point-like action (adding its definition by using a Lagrange multiplier), using the \textit{first prolongation} of the infinitesimal generator. Otherwise, one should use the generalised Euler-Lagrange equations for Lagrangian depending up to the second derivative in time, $\tfrac{\partial \mathcal{L}}{\partial\phi}-\tfrac{d}{dt}\tfrac{\partial \mathcal{L}}{\partial\dot{\phi}}+\tfrac{d^2}{dt^2}\tfrac{\partial \mathcal{L}}{\partial\ddot{\phi}}=0$. We will adopt the former approach, \textit{i.e.}, the Lagrange multiplier, as done in the previous section. This simplifies the resolution of the Noether identity, allowing us to use the relation $\ddot{\phi}={\dot{X}}/{\dot{\phi}}$.

After solving the Lagrange multiplier $\lambda = \left(\frac{3 \dot{a}}{a \dot{\phi}}-\tfrac{\dot{X}}{2 X \dot{\phi}}\right)G_3-\tfrac{\dot{\phi}}{a}\partial_{X}G_3+\partial_{\phi}G_3-\partial_{X}G_2$, the point-like Lagrangian turns into
\begin{align}\label{eq:braiding-lagr}
    \mathcal{L}=&-6a\dot{a}^2+a^3 G_{2}(\phi,X)-a^3\partial_{X}G_{2}(\phi,X)+a^3\,G_{3}(\phi,X)\left(\frac{\dot{X}}{\dot{\phi}}+3\frac{\dot{a}}{a}\dot{\phi}\right)\nn\\
    &+a^3\left(X-\frac{1}{2}\dot{\phi}^2\right)\left[\left(\frac{3 \dot{a}}{a \dot{\phi}}-\frac{\dot{X}}{2 X \dot{\phi}}\right)G_3(\phi,X)-\frac{\dot{\phi}}{a}\partial_{X}G_3(\phi,X)+\partial_{\phi}G_3(\phi,X)-\partial_{X}G_2(\phi,X)\right].
\end{align}
Then, in full generality, one obtains:
\begin{equation}\label{eq:braiding-gen}
    \xi(t)=\xi_1\,t+\xi_0\,,\quad\eta_{a}(a)=\frac{a}{3}\,\xi_1\,,\quad\eta_{\phi}=\eta_\phi(\phi)\,,\quad\eta_{X}=2X\left(\partial_{\phi}\eta_{\phi}-\xi_1\right)\,,
\end{equation}
with the following selected configurations,
\begin{align}\label{eq:braiding-eta}
    {\rm\left(\,I\,\right)}:\quad&\begin{cases}
        \vspace{-10pt}\\
        \xi_1\neq0\vspace{10pt}\\
        \eta_{\phi}(\phi)=\xi_1 \phi+\xi_0\vspace{10pt}\\
        G_{3}(\phi,X)=\dfrac{g_3(X)}{\xi_1\phi+\xi_0}\vspace{10pt}\\
        G_{2}(\phi,X)=\dfrac{g_2(X)}{(\xi_1\phi+\xi_0)^2}\vspace{3pt}
    \end{cases}\\[10pt]
    {\rm\left(\,II\,\right)}:\quad&\begin{cases}
        \vspace{-10pt}\\
        \eta_{\phi}(\phi)\neq\xi_1 \phi+\xi_0\vspace{10pt}\\
        G_{3}(\phi,X)=\exp \left(-\int _1^{\phi }\frac{\partial_{\varphi}\eta_{\phi}(\varphi)}{\eta_{\phi} (\varphi)}d\varphi\right) g_3\left(X \exp \left(\int _1^{\phi }\frac{2 \left(\xi_1-\partial_{\varphi}\eta_{\phi}(\varphi)\right)}{\eta_{\phi} (\varphi)}d\varphi\right)\right)\vspace{10pt}\\
         G_{2}(\phi,X)=\exp \left(\int _1^{\phi }-\frac{2 \xi_1}{\eta_{\phi} (\varphi)}d\varphi\right) g_2\left(X \exp \left(\int _1^{\phi }\frac{2 \left(\xi_1-\partial_{\varphi}\eta_{\phi} (\varphi)\right)}{\eta_{\phi} (\varphi)}d\varphi\right)\right)\\[7pt]
        \qquad\qquad\quad\,-2 X \exp \left(\int _1^{\phi }\frac{-2 \partial_{\varphi}\eta_{\phi} (\varphi)}{\eta_{\phi} (\varphi)}d\varphi\right) g_3\left(X \exp \left(\int _1^{\phi }\frac{2 \left(\xi_1-\partial_{\varphi}\eta_{\phi} (\varphi)\right)}{\eta_{\phi} (\varphi)}d\varphi\right)\right)\\[7pt]
        \qquad\qquad\quad\,\times\int _1^{\phi }\exp \left(\int _1^{\varphi_2}\frac{\partial_{\varphi_1}\eta_{\phi}(\varphi_1)}{\eta_{\phi} (\varphi_1)}d\varphi_1\right)\frac{ \partial_{\varphi_2}^2\eta_{\phi} (\varphi_2)}{\eta_{\phi} (\varphi_2)}d\varphi_2 \vspace{3pt}\vspace{3pt}
    \end{cases}\vspace{5pt}
\end{align}
where, $\xi_{0,1,2}$ are constants, and $g_{2,3}$ are arbitrary functions of $X$ times a factorised scalar field dependence (eventually constant). Notice that, in the case of internal symmetries the theory is manifestly shift-symmetric, \textit{i.e.}, $G_{i}=G_{i}(X)$.

\section{Horndeski gravity}\label{sec:horn}
\setcounter{equation}{0}

The Horndeski action reads as follows,
\begin{equation}
S= \int d^4 x \sqrt{-g} \, \left( { L}_2 + { L}_3+ { L}_4+ { L}_5 \right) , \label{Horndeskiaction}
\end{equation}
where, 
\begin{align}
{L}_2 =\,& G_2\left( \phi, X \right) \,,\nn\\ 
&\nn\\
{L}_3 =\,&- G_3\left( \phi, X \right) \Box \phi \,,\nn\\ 
&\nn\\
{L}_4 =\,& G_4\left( \phi, X \right) R +\partial_{X}G_{4} ( \phi, X ) 
\left[ \left( \Box \phi \right)^2 -\left( \nabla\nabla \phi \right)^2 
\right] \,, \nonumber\\ 
&\nn\\
{L}_5  =\,&  G_5\left( \phi, X \right) G_{ab}\nabla^a \nabla^b \phi -\frac{ 1}{6}\,\partial_{X}G_{5}(\phi,X) \left[ \left( \Box\phi \right)^3 -3\Box\phi \left( \nabla\nabla \phi \right)^2 +2\left( \nabla \nabla \phi \right)^3 \right],
\end{align}
where $(\nabla\nabla\phi)^2=\nabla_a\nabla_b\phi\nabla^a\nabla^b\phi$ and $(\nabla\nabla\phi)^3=\nabla_a\nabla_c\phi\nabla^a\nabla^b\phi\nabla^c\nabla_b\phi$.

Unlike the kinetic braiding function, the presence of $G_4$ and $G_5$ does not introduce an explicit dependence on second derivatives of the scalar field. All the factors having second derivatives can be integrated by parts yielding a first-order Lagrangian:
\begin{align}
    \mathcal{L}_{4}=\,&\,6 a^2\ddot{a} \,G_4(\phi,X)+6 a^2 \dot{a} \dot{\phi} \ddot{\phi}\,\partial_{X}G_{4}(\phi,X)+6 a \dot{a}^2 \left[\dot{\phi}^2 \partial_{X}G_4(\phi,X)+G_4(\phi,X)\right]\nn\\
    =\,&-6 a^2 \dot{a} \dot{\phi} \,\partial_{\phi}G_4(\phi ,X)+6 a \dot{a}^2 \left[\dot{\phi}^2 \partial_{X}G_4(\phi,X)-G_4(\phi,X)\right],\\
    &\nn\\
    \mathcal{L}_{5}=\,&\,3 a \dot{a}^2 \ddot{\phi}\left[\dot{\phi}^2 \,\partial_{X}G_{5}(\phi,X)+G_{5}(\phi,X)\right]+6 a \dot{a} \ddot{a}\dot{\phi} \,G_{5}(\phi,X)+\dot{a}^3 \dot{\phi} \left[\dot{\phi}^2 \,\partial_{X}G_{5}(\phi,X)+3 G_{5}(\phi,X)\right]\nn\\
    =\,&\,\dot{a}^2 \dot{\phi}^2 \left[\dot{a} \dot{\phi} \,\partial_{X}G_{5}(\phi,X)-3 a \,\partial_{\phi}G_5(\phi,X)\right].
\end{align}
Then, introducing the Lagrange multiplier, the point-like Lagrangian reads as follows,
\begin{align}\label{eq:horn-lagr}
    \mathcal{L}=\,&\, -6 a^2 \dot{a} \dot{\phi} \partial_{\phi}G_{4} +6 a \dot{a}^2 \left(2 X \partial_{X}G_{4} -G_{4} \right)+a^3 G_{2} +\frac{a^3 \dot{X} G_{3} }{\dot{\phi}}+3 a^2 \dot{a} \dot{\phi} G_{3} +2 X \dot{a}^2 \left(\dot{a} \dot{\phi} \partial_{X}G_{5} -3 a \partial_{\phi}G_{5} \right)\nn\\
    &+a^3 \left(X-\frac{1}{2} \dot{\phi}^2\right) \Bigg[-\partial_{X}G_{2} +\partial_{\phi}G_{3} -\frac{3 \dot{a} \dot{\phi} \partial_{X}G_{3} }{a}+\frac{\left(6 X \dot{a}-a \dot{X}\right) G_{3} }{2 a X \dot{\phi}}\nn\\
    &+\frac{6 \dot{a} \dot{\phi} \partial_{\phi X} G_{4} }{a}-\frac{6 \dot{a}^2 \partial_{X}G_{4} }{a^2}-\frac{12 X \dot{a}^2 \partial_{X}^2G_{4} }{a^2}-\frac{2 \dot{a}^3 \dot{\phi} \partial_{X}G_{5} }{a^3}-\frac{2 X \dot{a}^3 \dot{\phi} \partial_{X}^2G_{5} }{a^3}+\frac{6 \dot{a}^2 \partial_{\phi}G_{5} }{a^2}+\frac{6 X \dot{a}^2 \partial_{\phi X} G_{5} }{a^2}\Bigg].
\end{align}

From the Noether Symmetry Approach, one obtains the following configurations:
\begin{equation}\label{eq:horn-gen}
    \xi(t)=\xi_1\,t+\xi_2\,,\quad\eta_{a}(a)=\frac{a}{3}(\xi_1-\phi_0)\,,\quad\eta_{\phi}=\eta_{\phi}(\phi)\,,\quad\eta_{X}(\phi,X)=2X\left(\partial_{\phi}\eta_{\phi}-\xi_1\right)
\end{equation}
with,
\begin{align}
    {\rm\left(\,I\,\right)}:\quad&\begin{cases}\label{eq:horn-eta}
        \vspace{-10pt}\\
        \xi_1\neq0\vspace{10pt}\\
        \eta_{\phi}(\phi)=\xi_1 \phi+\xi_0\vspace{10pt}\\
        G_{5}(\phi,X)=\alpha\,\phi+ g_5(X)\left(\xi_1 \phi+\xi_0\right)^{\frac{\phi_0}{\xi_1}+1}\vspace{10pt}\\
        G_{4}(\phi,X)=\alpha X+\beta\sqrt{2X}+g_4(X)\left(\xi_1 \phi+\xi_0\right)^{\phi_0/\xi_1}\vspace{10pt}\\
        G_{3}(\phi,X)=g_3(X)\left(\xi_1 \phi+\xi_0\right)^{\frac{\phi_0}{\xi_1}-1}\vspace{10pt}\\
        G_{2}(\phi,X)=g_2(X)\left(\xi_1 \phi+\xi_0\right)^{\frac{\phi_0}{\xi_1}-2}\vspace{5pt}
    \end{cases}\\[10pt]
    {\rm\left(\,II\,\right)}:\quad&\begin{cases}\label{eq:horn}
        \vspace{-7pt}\\
        \eta_{\phi}(\phi)\neq\xi_1 \phi+\xi_0\vspace{10pt}\\
        G_{5}(\phi,X)=\alpha\,\phi+\exp \left(\int _1^{\phi }\frac{4 \xi_1+\phi_0-3 \partial_{\varphi}\eta_{\phi}(\varphi)}{\eta_{\phi} (\varphi)}d\varphi\right)\int _1^Xg_{5}\left(Z \exp \left(-\int _1^{\phi }-\frac{2 \left(\xi_1-\partial_{\varphi}\eta_{\phi}(\varphi)\right)}{\eta_{\phi} (\varphi)}d\varphi\right)\right) dZ\vspace{10pt}\\
        G_{4}(\phi,X)=\alpha X+\beta\sqrt{2X}+\int _1^{\phi }\exp\left({\int _1^{\varphi_2}\frac{\phi_0-\partial_{\varphi_1}\eta_{\phi}(\varphi_1)}{\eta_{\phi} (\varphi_1)}d\varphi_1}\right) g_{4}\left(X \exp \left(\int _1^{\varphi_2}\frac{2 \left(\xi_1-\partial_{\varphi_1}\eta_{\phi}(\varphi_1)\right)}{\eta_{\phi} (\varphi_1)}d\varphi_1\right)\right)d\varphi_2\vspace{10pt}\\
        G_3(\phi,X)=\exp \left(\int _1^{\phi }\frac{\phi_0-\partial_{\varphi}\eta_{\phi}(\varphi)}{\eta_{\phi} (\varphi)}d\varphi\right) g_3\left(X \exp \left(\int _1^{\phi }\frac{2 \left(\xi_1-\partial_{\varphi}\eta_{\phi}(\varphi)\right)}{\eta_{\phi} (\varphi)}d\varphi\right)\right)\vspace{10pt}\\
        G_{2}(\phi,X)=\exp \left(\int _1^{\phi }-\frac{2 \xi_1}{\eta_{\phi} (\varphi)}d\varphi\right) g_2\left(X \exp \left(\int _1^{\phi }\frac{2 \left(\xi_1-\partial_{\varphi}\eta_{\phi} (\varphi)\right)}{\eta_{\phi} (\varphi)}d\varphi\right)\right)\\[7pt]
        \qquad\qquad\quad\,-2 X \exp \left(\int _1^{\phi }\frac{-2 \partial_{\varphi}\eta_{\phi} (\varphi)}{\eta_{\phi} (\varphi)}d\varphi\right) g_3\left(X \exp \left(\int _1^{\phi }\frac{2 \left(\xi_1-\partial_{\varphi}\eta_{\phi} (\varphi)\right)}{\eta_{\phi} (\varphi)}d\varphi\right)\right)\\[7pt]
        \qquad\qquad\quad\,\times\int _1^{\phi }\exp \left(\int _1^{\varphi_2}\frac{\partial_{\varphi_1}\eta_{\phi}(\varphi_1)}{\eta_{\phi} (\varphi_1)}d\varphi_1\right)\frac{ \partial_{\varphi_2}^2\eta_{\phi} (\varphi_2)}{\eta_{\phi} (\varphi_2)}d\varphi_2 \vspace{3pt}\vspace{5pt}
    \end{cases}
\end{align}
where, $\xi_{0,1,2}$, $\alpha$, $\beta$, and $\phi_0$ are constants, and $g_{2,3,4,5}$ are arbitrary functions of $X$ times a factorised scalar field dependence (eventually constant). 

The constants $\alpha$ and $\beta$ can be set to zero. Indeed, it is straightforward to notice that $L_4$ vanishes in the case of $G_{4}(\phi,X)=\beta\sqrt{2X}$. This represents a spurious solution of the Noether approach. Moreover, it is well-known that in the case of $G_{5}=G_{5}(\phi)$, one can reabsorb the $G_5$ dependence by redefining $G_{2,3,4}$ as follows,
\begin{eqnarray}\label{eq:trans_law}
G_{2}  \to  G_{2}-2X^2\partial^{3}_{\phi}G_{5} \,,\,\quad G_{3} \to  G_{3} -3X\partial^{2}_{\phi}G_{5} \,,\,\quad G_4 \to  G_{4}-X\partial_{\phi}G_{5} \,,
\end{eqnarray}
because it turns out that
\begin{eqnarray}
L_{5} &=& G_5G_{ab}\nabla^{a}\nabla^{b}\phi\simeq 
-\partial_{\phi}G_{5}G_{ab}\nabla^{a}\phi\nabla^{b}\phi = -\partial_{\phi}G_{5}R_{ab}\nabla^{a}\phi\nabla^{b}\phi
-X\partial_{\phi}G_{5}R \nonumber\\
&\nn\\
&=& \partial_{\phi}G_{5}(\nabla_{a} 
\Box\phi-\nabla_{b}\nabla_{a}\nabla^{b}\phi) 
\nabla^{a}\phi-X\partial_{\phi}G_{5}R \nonumber\\
&\nn\\
& \simeq & - \partial^{2}_{\phi}G_{5}(-2X\Box\phi-\nabla_{a}\nabla_{b}\phi 
\nabla^{a}\phi\nabla^{b}\phi)  -\partial_{\phi}G_{5}
\left[\Box\phi^2-(\nabla_{a} 
\nabla_{b}\phi)^2\right]-X\partial_{\phi}G_{5}R \nn\\
&\nn\\
& \simeq & 3X\,\partial^{2}_{\phi}G_{5}\,\Box\phi-2X^2\partial^{3}_{\phi}G_{5}  -\partial_{\phi}G_{5}
\left[\Box\phi^2-(\nabla_{a} 
\nabla_{b}\phi)^2\right]-X\partial_{\phi}G_{5}R\,,
\end{eqnarray}
where $\simeq$ means equality up to a total divergence. In particular, from Eq.~\eqref{eq:trans_law}, it is possible to notice that, $G_{5}=\alpha\phi$ is equivalent to $G_{4}=-\alpha X$. For this reason, it is possible to set $\alpha=0$ in Eqs.~\eqref{eq:horn-eta} and~\eqref{eq:horn}, without losing generality. Equivalently, one can verify that $\mathcal{L}_{4}+\mathcal{L}_{5}=0$ if $G_4=\alpha\, X$ and $G_{5}=\alpha\,\phi$.

The Noether symmetries classification of viable Horndeski gravity, $\partial_{X}G_4=0$ and $G_5=0$, can be obtained by considering $g_4\to\mbox{const}$ and $g_5\to0$. Then, in the case of external symmetries~\eqref{eq:horn-eta} the parameter $\phi_0$ is associated with the presence of a non-minimal coupling $G_{4}=G_{4}(\phi)$, while, in the case of external symmetry, $\phi_0$ represents the breaking parameter of the shift-symmetry.

From the performed analysis of Noether symmetries, it is possible to notice that, in the more general scalar-tensor theories, the selected models can be written such that the $\phi$ and $X$ dependence is factorised:
\begin{equation}\label{eq:general-noether}
    G_{i}(\phi,X)=h_{i}(\phi)\,g_{i}(X)\qquad
    \begin{cases}
        \vspace{-7pt}\\
        h_{i}(\phi)={\partial_{\phi}^{(6-i)}}(\xi_1\phi+\xi_0)^{2+{\phi_0}/{\xi_1}}\,, \quad \mbox{if}\quad\xi_{1}\neq0\vspace{10pt}\\
        h_{i}(\phi)=\exp\left(\tfrac{\phi_0}{\xi_0}\phi\right)\,, \quad\mbox{if}\quad \xi_{1}=0\vspace{7pt}
    \end{cases}
\end{equation}
In particular, for the symmetries having $\xi_{1}\neq0$, the scalar field can be redefined so that $\eta_{\phi}=\xi_1\phi+\xi_0$; then, it turns out that $h(\phi)=(\xi_1\phi+\xi_0)^{2+{\phi_0}/{\xi_1}}$. In the case of internal symmetries, which are characterised by $\xi_1=0$, the scalar field can be redefined so that $\eta_{\phi}=\xi_0$; then, it turns out that $h(\phi)=\exp\left(\tfrac{\phi_0}{\xi_0}\phi\right)$. Finally, notice that the shift-symmetric class is characterised by $G_{i}=G_{i}(X)$, corresponding to $\xi_1=0=\phi_0$. Less general subclasses can be obtained from the above system by manually setting the functions $g_{i}$ to 1 and/or $\phi_{0}=0$, except for the linear minimally coupled scalar field and the first-generation scalar-tensor theory, which are characterised by a more complex substructure.

\section{Particular cases}\label{sec:models}
\setcounter{equation}{0}

The Noether Symmetry Approach almost fully determines the $\phi$ dependence, while the $g_{i}(X)$ functions are unconstrained. Then, each possible set $\{g_{i}\}_{i=2,3,4,5}$ corresponds to a particular model contained in Horndeski gravity. Therefore, models admitting Noether symmetries are characterised by the $X$ dependence of the $g_{i}$ functions. Once the model has been selected, \textit{i.e.}, the $X$ dependence, the request of a Noether symmetry sets the remaining free functions of $\phi$.
To clarify this point, let us discuss the cases of two different theories assuming the existence of a Noether symmetry.

\subsection{Non-minimal coupling to the Gauss-Bonnet term}

The Gauss-Bonnet topological invariant is a combination of second-order curvature invariants defined as
\begin{equation}
    \mathcal{G}=R^{2}-4 R_{\mu\nu}R^{\mu\nu}+R_{\mu\nu\rho\sigma}R^{\mu\nu\rho\sigma}\,.
\end{equation}
The corresponding action is a topological term, which can be written as a total derivative (\textit{i.e.}, a boundary term), representing the four-dimensional case of the Chern-Gauss-Bonnet theorem~\cite{Nakahara:1990th}. It states that the Euler characteristic of an oriented closed even-dimensional Riemannian manifold is equal to the integral of a certain polynomial of its curvature~\cite{Nojiri:2005jg, Nojiri:2005vv, Nojiri:2017ncd}. 
Due to its topological nature, the Gauss-Bonnet invariant is often considered a tool to reduce the dynamics. However, to make its contribution not trivial in four dimensions, it is usually either coupled to a dynamical scalar field or included in the Einstein–Hilbert action\footnote{One of the concerns about the presence of $\mathcal{G}$ into the gravitational action is the impossibility of imposing gravitational waves travelling at the speed of light in a covariant way. This represents another open issue in modified theories of gravity. However, this topic is beyond the scope of this article.} as a generic function $f(\mathcal{G})$.
Thus, as the scalar curvature is predominant at local scales, the Gauss-Bonnet correction might provide corrections at cosmological scales.
Let us take into account the former case,
\begin{equation}
    S_{\mathcal{G}}=\int{d^4x\sqrt{-g}\, h(\phi)\,\mathcal{G}}\,.
\end{equation}
In this regard, it is well-known that Horndeski's theory can reproduce the non-minimal coupling to the Gauss-Bonnet term, since it represents the most general theory in four dimensions of $\phi$, $g_{ab}$, and their derivatives, giving the second-order field equations~\cite{Kobayashi:2019hrl}. The corresponding Horndeski contributions are as follows:
\begin{align}
    G_{5}(\phi,X)=\,&-4\,\partial_{\phi}h(\phi)\ln{X}\,,\\
    G_{4}(\phi,X)=\,&\,4\,\partial_{\phi}^2 h(\phi)\,X\left(2-\ln{X}\right),\\
    G_{3}(\phi,X)=\,&\,4\,\partial_{\phi}^{3}h(\phi)\,X\left(7-3\ln{X}\right),\\
    G_{2}(\phi,X)=\,&\,8\,\partial_{\phi}^{4}h(\phi)\,X^2\left(3-\ln{X}\right).
\end{align}
The easiest way to prove this is by directly comparing the equations yielding from the variation with respect to the metric tensor on the spatially flat FLRW background.

It is straightforward to see that the above set of functions $G_{i}$ is compatible with any of the selected Noether symmetries; not only is the hierarchical derivative dependence the same as the first model, but it is also the same as the function redefinition to absorb $G_{5}=G_{5}(\phi)$. Then, depending on the Noether symmetry, $h(\phi)=(\xi_1\phi+\xi_0)^{2+{\phi_0}/{\xi_1}}$ or $h(\phi)=(\xi_1\phi+\xi_0)^{2+{\phi_0}/{\xi_1}}$ or $h(\phi)=\exp\left(\tfrac{\phi_0}{\xi_0}\phi\right)$.

It is possible to analyse the non-minimal coupling to the Gauss-Bonnet terms in the Noether framework. Consistently with the Noether symmetries, the three simplest actions that one can take into account are of the following form
\begin{equation}
    S=\int{d^4x\sqrt{-g}\left[f(\phi)R+v(\phi)\left(\omega_{0}\,X-V_{0}\right)+h(\phi)\,\mathcal{G}\right]}\,,
\end{equation}
where,
\begin{equation}
    \begin{cases}
        \vspace{-10pt}\\
        h=c_{h}\left(\xi_1\phi+\xi_0\right)^{2+{\phi_0}/{\xi_1}}\,,\enspace f=c_{f}\,\partial_{\phi}^{2}h\,,\enspace  v=c_{v}\,\partial_{\phi}^{4}h\,, \quad\mbox{if}\quad\left\{\xi=\xi_1\,t,\,\eta_{a}=\dfrac{a}{3}(\xi_1-\phi_0),\,\eta_{\phi}=\xi_1\,\phi+\xi_0\right\}\vspace{10pt}\\
        h=c_{h}\exp\left(\tfrac{\phi_0}{\xi_0}\phi\right),\enspace f=c_{f}\exp\left(\tfrac{\phi_0}{\xi_0}\phi\right), \enspace v=c_{v}\exp\left(\tfrac{\phi_0}{\xi_0}\phi\right),\quad\mbox{if}\quad \left\{\xi=0,\,\eta_{a}=-\dfrac{a}{3}\,\phi_0,\,\eta_{\phi}=\xi_0\right\}\vspace{10pt}\\
        h=c_{h}\,\phi\,,\enspace f=c_{f}\,,\enspace v=c_{v}\,,\quad\mbox{if}\quad \left\{\xi=0,\,\eta_{a}=0,\,\eta_{\phi}=\xi_0\right\}\vspace{5pt}
    \end{cases}
\end{equation}
The corresponding Horndeski model is
\begin{align}
    G_{5}(\phi,X)=\,&-4\,\partial_{\phi}h(\phi)\ln{X}\,,\\
    G_{4}(\phi,X)=\,&f(\phi)+4\,\partial_{\phi}^2 h(\phi)\,X\left(2-\ln{X}\right),\\
    G_{3}(\phi,X)=\,&\,4\,\partial_{\phi}^{3}h(\phi)\,X\left(7-3\ln{X}\right),\\
    G_{2}(\phi,X)=\,&v(\phi)\left(\omega_{0}\,X-V_{0}\right)+8\,\partial_{\phi}^{4}h(\phi)\,X^2\left(3-\ln{X}\right).
\end{align}
Field equations together with the Hamiltonian constraint (the first Friedmann equation) read, respectively,
%\footnote{Notice that, since we have a specific expression for $G_3$, we can also choose not to use the Lagrange multiplier and instead integrate by parts to eliminate the second derivative of the scalar field inside the point-like Lagrangian.}
%
\begin{align}
    \frac{\partial\mathcal{L}}{\partial a}-\frac{d}{dt}\frac{\partial\mathcal{L}}{\partial \dot{a}}=0 \quad\longrightarrow\quad&3 X \left(4 \partial^{2}_{\phi}f+\omega _0 v\right)+12H\dot{\phi} \left(\partial_{\phi}f+4\dot{H} \partial_{\phi}h+4 H^2 \partial_{\phi}h\right)\nn\\
    &+6\ddot{\phi} \left( \partial_{\phi}f+4 H^2 \partial_{\phi}h\right)+6 H^2 \left(3 f+8 X \partial_{\phi}^2h\right)+12 f \dot{H}-3 V_0 v=0\,,\\
    &\nn\\
    \frac{\partial\mathcal{L}}{\partial \phi}-\frac{d}{dt}\frac{\partial\mathcal{L}}{\partial \dot{\phi}}=0 \quad\longrightarrow\quad&\dot{\phi} \left[6\dot{H} \left( \partial_{\phi}f+4 H^2 \partial_{\phi}h\right)+12 H^2( \partial_{\phi}f+2 H^2 \partial_{\phi}h)-\partial_{\phi}v \left(\omega _0 X+V_0\right)\right]\nn\\
    &-6 \omega _0 H X v-\omega _0 v \dot{\phi} \ddot{\phi}=0\,,\\
    &\nn\\
    \frac{\partial\mathcal{L}}{\partial X}-\frac{d}{dt}\frac{\partial\mathcal{L}}{\partial \dot{X}}=0 \quad\longrightarrow\quad&X=\frac{1}{2}\dot{\phi}^2\,,\\
    &\nn\\
    \dot{a}\frac{\partial\mathcal{L}}{\partial \dot{a}}+\dot{\phi}\frac{\partial\mathcal{L}}{\partial \dot{\phi}}+\dot{X}\frac{\partial\mathcal{L}}{\partial \dot{X}}-\mathcal{L}=0 \quad\longrightarrow\quad&\dot{\phi} \left[v  \left(\omega _0 X+V_0\right)-6 H^2 f \right]-12 H X( \partial_{\phi}f +4 H^2  \partial_{\phi}h) =0\,,
\end{align}
and, taking into account the above energy constraint, the conserved scalar current turns out to be
\begin{align}
    &\mathcal{J}=\zeta-\eta_{a}\frac{\partial\mathcal{L}}{\partial \dot{a}}-\eta_{\phi}\frac{\partial\mathcal{L}}{\partial \dot{\phi}}-2X\left(\partial_{\phi}\eta_{\phi}-\xi_1\right)\frac{\partial\mathcal{L}}{\partial \dot{X}}\nn\\
    &\Rightarrow\quad\dot{\phi} \,\Big\{2 (\xi_1-\phi_0) \partial_\phi f+8 H^2 (\xi_1-\phi_0) \partial_{\phi}h-\omega _0 \eta_{\phi} v\nn\\
    &\qquad+4 X \left[\eta_{\phi} \partial^{4}_{\phi}h (\ln{X}-3)+\partial^{3}_{\phi}h \left(4 \xi_1+\partial_{\phi}\eta_{\phi} (3 \ln{X}-7)-(2 \xi_1+\phi_0) \ln{X}+3 \phi_0\right)\right]\!\Big\}\nn\\
    &\qquad+2H \left[3 \eta_{\phi} \partial_{\phi}f+2 (\xi_1-\phi_0) f\right]+8 H^3 \eta_{\phi} \partial_{\phi}h=\frac{\ell}{a^3}\,.
\end{align}
where, $\ell$ is a constant introduced for practical reasons, by redefining $\zeta$. Then, it is possible to find exact solutions as shown in~\cite{Capozziello:2018gms, Bajardi:2020xfj, Bajardi:2020osh, Bajardi:2022ypn}.

\subsection{Extended cuscuton model}

The extended cuscuton model~\cite{Iyonaga:2018vnu, Iyonaga:2020bmm} is a generalised formulation of the cuscuton field~\cite{Afshordi:2006ad, Afshordi:2007yx, Afshordi:2009tt}, which is not dynamic at the background and perturbation levels. The action of the model corresponds to the following choice of the Horndeski functions,
\begin{align}
    &G_4(\phi)=f_4(\phi)\,,\\
    &G_3(\phi,X)=\left(\dfrac{1}{2}f_3(\phi)+\partial_{\phi}f_{4,\phi}\right)\ln{X}\,,\\
    &G_2(\phi,X)=f_1(\phi)+f_2(\phi)\sqrt{2X}-\left(2\partial_{\phi}f_{3}(\phi)+4\partial^{2}_{\phi}f_{4}(\phi)+\frac{3{f_3}(\phi)^2}{4f_4(\phi)}\right)X\nn\\
    &\qquad\qquad\quad+2\left(\frac{1}{2}\partial_{\phi}f_{3}(\phi)+\partial_{\phi}^2f_{4}(\phi)\right)X\ln{X}\,.\\
\end{align}

As it is possible to notice for the above equations, the model has an explicit $X$ dependence, while the $\phi$ dependence is parameterised by the presence of the function $f_i$. However, imposing the existence of Noether symmetries, our previous analysis provides a criterion to select them according to the following scheme:
\begin{equation}
    \begin{cases}
        f_{4}=c_4\,(\xi_1\phi+\xi_0)^{\phi_0/\xi_0}\,,\enspace f_{3}=c_3\,(\xi_1\phi+\xi_0)^{\frac{\phi_0}{\xi_0}-1}\,,\enspace f_{1,2}=c_{1,2}\,(\xi_1\phi+\xi_0)^{\frac{\phi_0}{\xi_0}-2}\quad\mbox{if}\quad\xi\neq0\vspace{10pt}\\
        f_{i}=c_i\exp\left(\tfrac{\phi_0}{\xi_0}\phi\right)\,, \quad \mbox{if}\quad\xi=0\vspace{7pt}
    \end{cases}
\end{equation}
where $c_i$ are generic constants. Then, one can write down field equations and the energy condition,
\begin{align}
    \frac{\partial\mathcal{L}}{\partial a}-\frac{d}{dt}\frac{\partial\mathcal{L}}{\partial \dot{a}}=0 \quad\longrightarrow\quad& f_{1}+  \sqrt{2X} f_{2}-2 X \partial_{\phi}f_{3}-\frac{3f_{3}^2}{4 f_{4}}X- f_{3} \ddot{\phi}\nn\\
    &+4 H \dot{\phi} \partial_{\phi}f_{4}+2(2 \dot{H}+3 H^2)f_{4}=0\,,\\
    &\nn\\
    \frac{\partial\mathcal{L}}{\partial \phi}-\frac{d}{dt}\frac{\partial\mathcal{L}}{\partial \dot{\phi}}=0 \quad\longrightarrow\quad&\dot{\phi} \left[\frac{\partial_{\phi}f_{1}}{3}-X\frac{f_{3} }{4 f_{4}^2}\left(f_{3} \partial_{\phi}f_{4}-2 f_{4} \partial_{\phi}f_{3}\right)-H^2 \left(3 f_{3}+2 \partial_{\phi}f_{4}\right)- f_{3} \dot{H}\right]\nn\\
    &+H \left(3X\frac{f_{3}^2}{2 f_{4}}-  \sqrt{2X} f_{2}\right)+\frac{ f_{3}^2 }{4 f_{4}}\dot{\phi} \ddot{\phi}=0\,,\\
    &\nn\\
    \frac{\partial\mathcal{L}}{\partial X}-\frac{d}{dt}\frac{\partial\mathcal{L}}{\partial \dot{X}}=0 \quad\longrightarrow\quad&X=\frac{1}{2}\dot{\phi}^2\,,\\
    &\nn\\
    \dot{a}\frac{\partial\mathcal{L}}{\partial \dot{a}}+\dot{\phi}\frac{\partial\mathcal{L}}{\partial \dot{\phi}}+\dot{X}\frac{\partial\mathcal{L}}{\partial \dot{X}}-\mathcal{L}=0 \quad\longrightarrow\quad&\dot{\phi} \left(\frac{f_{1}}{3}+X\frac{f_{3}^2}{4 f_{4}}+2 H^2 f_{4}\right)-2 H X f_{3}=0\,,
\end{align}
while the \textit{on-shell} conserved scalar current is
\begin{align}
    &\mathcal{J}=\zeta-\eta_{a}\frac{\partial\mathcal{L}}{\partial \dot{a}}-\eta_{\phi}\frac{\partial\mathcal{L}}{\partial \dot{\phi}}-2X\left(\partial_{\phi}\eta_{\phi}-\xi_1\right)\frac{\partial\mathcal{L}}{\partial \dot{X}}\nn\\
    &\Rightarrow\quad \dot{\phi} \Bigg[-\frac{4 f_{2} \eta_{\phi} }{\sqrt{2X}}-2 \eta_{\phi}  (\ln{X}-2) \left(\partial_{\phi}f_{3}+2 \partial_{\phi}\partial_{\phi}f_{4}\right)+\frac{3 f_{3}^2 \eta_{\phi} }{f_{4}}+2 f_{3} \ln{X} \left(\phi_{0}-\partial_{\phi}\eta_{\phi}\right)\nn\\
    &\qquad+4 \partial_{\phi}f_{4} \left(2 \xi_{1}-\partial_{\phi}\eta_{\phi} \ln{X}+\phi_{0} \ln{X}-2 \phi_{0}\right)\Bigg]+4H [4 (\xi_{1}-\phi_{0}) f_{4}-3 f_{3} \eta_{\phi} ]=\frac{\ell}{a^3}\,,
\end{align}
where, $\ell$ is a constant introduced for practical reasons, by redefining $\zeta$.

The Noether symmetry allows us to find all general solutions for the extended cuscuton model. In particular, in the case of the external symmetries, it turns out that, for $\ell=0$,
\begin{equation}
    H=\left(\dot{\phi}\frac{c_{3} }{4c_{4}}-\frac{ c_{2}}{3 c_{3}-4 c_{4}( \xi_{1}-\phi_0)}{}\right) (\xi_{0}+\xi_{1} \phi)^{-1}\,,
\end{equation}
with $c_1= -{6 c_2^2 c_4}{(3 c_3-4 c_4 \xi_1+4 c_4 \phi_0)^{-2}}$, while, for $\ell\neq0$, it yields $c_1=c_2=0$, $c_3=\frac{4}{3} c_4 (\xi_1-\phi_0)$, and
\begin{equation}
    H= \frac{(\xi_1-\phi_0) }{3 (\xi_0+\xi_1 \phi )}\dot{\phi}\quad\Rightarrow\quad a(t)= c_{a}\, (\xi_1 \,\phi+\xi_0)^{\frac{\xi_1-\phi_0}{3 \xi_1}}\,,
\end{equation}
where $c_a$ is a constant depending on the other free parameters.

In the case of internal symmetries, the exact solutions are, for $\ell=0$,
\begin{equation}
    H=\frac{c_3}{4 c_4}\dot{\phi}-\frac{c_2 \xi_0}{3 c_3 \xi_0+4 c_4 \phi_0}\quad\Rightarrow\quad a(t)=c_{a}\exp{\left(\frac{c_3}{4 c_4}\,\phi -\frac{c_2 \xi_0 }{3 c_3 \xi_0+4 c_4 \phi_0}\,t\right)}\,,
\end{equation}
with $c_1= -{6 c_4 c_2^2 \xi_0^2}{\left(3 c_3 \xi_0 + 4 c_4 \phi_0 \right)^{-2}}$, while, for $\ell\neq0$, it provides $c_1=c_2=0$, $c_3= -c_4 \frac{4}{3} \frac{\phi_0}{ \xi_0}$, and
\begin{equation}
    H=-\frac{\phi_0}{3 \xi_0}\dot{\phi}\quad\Rightarrow\quad a(t)=c_{a}\exp\left({-\frac{1}{3}\frac{\phi_0 }{ \xi_0}\phi}\right)
\end{equation}
where $c_a$ is a constant depending on the other free parameters.

\section{Noether symmetries with matter}\label{sec:matter}
\setcounter{equation}{0}

So far, we have conducted our analysis neglecting the presence of the matter. Indeed, the modifications to GR field equations usually aim to describe a different behaviour of gravity at early or late cosmic time, or in correspondence with very high/low-energy scales when the standard matter contribution can be left out of the treatment. In this framework, modified GR is often used to obtain a dynamical formulation of dark energy, instead of the cosmological constant~\cite{DESI:2024uvr}.
However, generally, one has to deal with situations where the matter content cannot be neglected since it plays a crucial role. For instance, this is the case of several cosmological tests on modified theories of gravity, or, in general, observational cosmology~\cite{Koyama:2015vza,Nesseris:2017vor,Frusciante:2019xia}.
Therefore, it is necessary to include the matter in this analysis.
Let us consider a point-like Lagrangian accounting also the presence of the matter,
\begin{equation}
    \mathcal{L}=\mathcal{L}^{\rm(g)}+\mathcal{L}^{\rm(m)}\,,
\end{equation}
where, $\mathcal{L}^{\rm(g)}$ represents a general scalar-tensor theory describing the gravitational part, and $\mathcal{L}^{\rm(m)}$ is associated with the matter. In cosmology, the different species of standard matter are usually described by linear barotropic equations of state $P_{i}=w_{i}\rho_{i}$, where $P_{i}$ and $\rho_{i}$ are the isotropic pressure and the energy density, respectively, $w_{i}$ is the barotropic coefficient, and $i$ labels the different species such as radiation ($w=1/3$), dust ($w=0$), etc. However, sometimes more complex equations of state are also used to consider interactions or to model exotic dark matter (not simply as a dust fluid). Therefore, there is no univocal way to take into account the matter content, and its description changes depending on the cosmic era being considered and the complexity of the model. The linear barotropic equation of state is the simplest way to describe the different species that fill the universe.

As shown in the previous sections, the Noether Symmetry Approach is a powerful tool to obtain functional forms of $\mathcal{L}^{\rm(g)}$ under the \textit{caveat} that our theory possesses a Noether symmetry. However, it is an exact mathematical computation that can be affected by the way $\mathcal{L}^{(m)}$ is modelled. An explicit example is the case of matter content characterised by a linear barotropic equation of state. It can be included in the point-like Lagrangian by considering $\mathcal{L}^{\rm(m)}=\rho_{0}\,a^{-3w}$, with $\rho_0$ constant. Including this contribution, it is possible to verify that symmetries strictly survive in correspondence with $\xi=0$ and $w=0$: only internal symmetries admit the presence of (dust) matter. Therefore, the matter Lagrangian can strongly constrain the Noether Symmetry Approach. A possible solution to overcome this problem is to consider an additional constraint equation that removes the presence of the matter from the Noether identity, ensuring the existence of symmetries, and modifying the usual approach (as done in~\cite{Capozziello:2008ch}). However, one could also accept that the matter content breaks the Noether symmetry, characterising the theory only in vacuum.

In this section, we propose an alternative treatment that allows us to keep all the Noether classifications obtained so far, safely including the presence of matter. It is based on an additional scalar field $\psi$ describing the matter content, minimally coupled to gravity: $\mathcal{L}^{\rm(m)}=\mathcal{L}^{\rm(m)}[g_{ab},\psi,\dot{\psi}]$. For simplicity, let us consider the linear case
\begin{equation}\label{eq:matter_lag}
    \mathcal{L}^{\rm(m)}=a^{3}\left(\frac{1}{2}\dot{\psi}^2-U\right),
\end{equation}
where $U=U(\psi)$ is a general potential of the scalar field (whose form is constrained by the existence of Noether symmetries).

Using an additional scalar field to effectively describe the matter content is a reasonable choice. Indeed, in this way, one has a general effective description, and a simplified parameterization can be introduced \textit{a posteriori} in the cosmological analysis.
Then, it turns out $\sum_{i}\rho_{i}=\tfrac{1}{2}\dot{\psi}^2+U$ and $\sum_{i}P_{i}=\tfrac{1}{2}\dot{\psi}^2-U$, or equivalently, $\dot{\psi}^2=\sum_{i}(\rho_{i}+P_{i})$ and $U=\tfrac{1}{2}\sum_{i}(\rho_i-P_i)$. In doing this, it is possible to verify that the Noether classification for scalar-tensor theories is unchanged: the selected functional forms of the action are the same as in our previous analysis; the only differences are the symmetries, {\it i.e.} the components of the infinitesimal generators of the symmetries (depending on $\psi$ also, in general). It represents a formal proof of the validity of the Noether Symmetry Approach in the presence of matter.
Perhaps the simplest way to demonstrate this is by assuming the functional forms selected by the Noether Symmetry Approach and verifying the existence of the symmetries given by the (non-vanishing) infinitesimal generator, 
\begin{equation}
    \chi=\xi(t,a,\phi,\psi)\partial_{t}+\eta_{a}(t,a,\phi,\psi)\partial_{a}+\eta_{\phi}(t,a,\phi,\psi)\partial_{\phi}+\eta_{\psi}(t,a,\phi,\psi)\partial_{\psi}\,.
\end{equation}
Then, the above components of the infinitesimal generator given by the Noether Symmetry Approach for Horndeski gravity turn out to be
\begin{equation}\label{eq:gen_matt}
        \xi(t)=\xi_1 t+\xi_2\,,\quad \eta_{a}(a,\phi)=\frac{a}{3}\left(\xi_1-\phi_0\right)\,,\quad \eta_{\psi}(\psi)=\frac{1}{2}(\phi_0\,\psi+\xi_3)\,,
\end{equation}
while the potential $U$ must satisfy the following differential equation,
\begin{equation}\label{eq:potential}
    (\phi_0-2 \xi_1) U(\psi )-\frac{1}{2}  (\psi \, \phi_0+ \xi_3)U'(\psi )=0\,,
\end{equation}
selecting $U=U_0\, (\phi_0\,\psi +\xi_3 )^{2-\frac{4 \xi_1}{\phi_0}}$ for $\phi_0\neq0$, or $U=U_{0}\exp\left(-\tfrac{4\xi_1}{\xi_3}\psi\right)$ for $\phi_{0}=0$, together with the same classification of Sec.~\ref{sec:horn}. However, Eq.~\eqref{eq:matter_lag} represents the simplest Lagrangian to consider. Alternatively, it is possible to leave unspecified the Lagrangian $\mathcal{L}^{\rm (m)}$ of the matter sector, or consider multiple scalar fields~\cite{Miranda:2022uyk}. For instance, a generalisation of the previous case is given by the following multi-field Lagrangian,
\begin{equation}
    \mathcal{L}^{\rm (m)}=a^{3}\left(\frac{1}{2}\sum_{I=1}^{N}\dot{\psi}^2_{I}-U(\psi_J)\right)\,,
\end{equation}
where the potential $U$ depends in general on the multiple $\psi_{J}=\{\psi_1,\dots,\psi_N\}$. Consequently, one obtains an analogous result of Eq.~\eqref{eq:gen_matt} for each matter field,
\begin{equation}\label{eq:gen_matt_multi}
        \xi(t)=\xi_1 t+\xi_2\,,\quad \eta_{a}(a,\phi)=\frac{a}{3}\left(\xi_1-\phi_0\right)\,,\quad \eta_{\psi_I}(\psi_I)=\frac{1}{2}(\phi_0\,\psi_I+\xi_{3,I})\,,
\end{equation}
while Eq.~\eqref{eq:potential} turns into
\begin{equation}
    (\phi_0-2 \xi_1) U-\frac{1}{2}\sum_{I=1}^{N}  (\psi_{I} \, \phi_0+\xi_{3,I})\partial_{I}U=0\,,
\end{equation}
corresponding to $U=U_0\, (\phi_0\,\psi_{J} +\xi_{3,J} )^{2-\frac{4 \xi_1}{\phi_0}}F(\eta_{\psi_I}/\eta_{\psi_J})$ for $\phi_0\neq0$, or $U=U_{0}\exp\left(-\tfrac{4\xi_1}{\xi_{3,J}}\psi_{J}\right)F(\psi_{I}-\psi_{J}\,\xi_{3,I}/\xi_{3,J})$ for $\phi_{0}=0$, where $F$ is a generic $(N-1)-$dimensional function, with $I\neq J$.

Discussing all possible matter Lagrangians goes beyond the scope of this work. The main result of this section is that introducing a matter field (or multiple fields) leaves the Noether symmetries of Horndeski gravity unchanged compared to the vacuum case.

\section{Discussion and Conclusions}\label{sec:disc}
\setcounter{equation}{0}

Noether symmetries represent a powerful tool for simplifying and solving dynamical systems.
Applying the Noether Symmetry Approach in cosmology constitutes a method to select the functional form of effective Lagrangians. In addition, the presence of a conserved scalar charge associated with the symmetry reduces the dynamics, helping to find exact cosmological solutions.

We focused our analysis on Horndeski gravity and its subclasses, providing a general classification. Reversing the usual Noether theorem and assuming the invariance under Noether point symmetries, we selected the functional forms of $G_{i}$, the Horndeski functions. This result is achieved by using a Lagrange multiplier to treat the kinetic term as a new variable for the system. The Lagrange multiplier allows us to keep the braiding function $G_3$ general and turns the point-like Lagrangian into a Lagrangian of the first order in time derivatives. This is because the point-like Lagrangian depends on the second derivative of the scalar field which cannot be removed by integrating by parts due to the presence of a (general) braiding term, $G_{3}$. However, once the braiding function has an explicit $X$ dependence, we can always transform $\mathcal{L}_3$ into a point-like Lagrangian of the first order. For this reason, the equations of motion are associated with second-order Euler-Lagrange equations. Then, an equivalent alternative approach is to consider the \textit{second prolongation} of the infinitesimal generator of the Noether symmetry. 

From the Eq.~\eqref{eq:general-noether} it is possible to see that the Noether symmetry almost fully determines the dependence of the functions on $\phi$, while the $X$ dependence is factorised in unconstrained $g_{i}(X)$ functions. This means that models admitting Noether symmetries are determined by the $X$ dependence of the $G_{i}$ functions. Once the model has been selected, \textit{i.e.}, the $X$ dependence is fixed, requiring the existence of a Noether symmetry sets the $\phi$ dependence. To highlight this aspect, we discussed the case of non-minimal coupling to the Gauss-Bonnet term and the extended cuscuton model.

The general $\phi$-dependence of the selected Horndeski functions is mainly characterised by two free parameters, $\xi_1$ and $\phi_0$. The former is associated with transformations of the coordinate time (\textit{i.e.}, $\xi_1\neq0$ external symmetry, $\xi_1=0$ internal symmetry). The latter is connected with transformations of the scale factor. However, the analysis of Eqs.~\eqref{eq:horn-eta} and~\eqref{eq:general-noether} provides an additional interpretation of the parameter $\phi_0$. Restricting to viable subclasses of Horndeski gravity, $G_{4}=G_{4}(\phi)$ and $G_{5}(\phi,X)=0$, in the case of external symmetries, a non-vanishing $\phi_0$ is related with the presence of a non-minimal coupling $G_{4}$, while, in the case of internal symmetry, it represents the breaking parameter of the shift-symmetry.

Finally, we extended our analysis by taking into account a point-like Lagrangian describing the {matter} sector.
Since the Noether Symmetry Approach is an exact mathematical computation, the classification obtained in the vacuum can strongly be affected. Indeed, parameterising the matter Lagrangian as $\mathcal{L}^{\rm(m)}=\rho_{0}\,a^{-3w}$ describing a single linear barotropic perfect fluid, the only symmetries surviving for Horndeski gravity are in correspondence with $\xi_1=0$ and $w=0$. Moreover, since there is no unique and general parameterization of the matter Lagrangian, but it depends on the particular cosmic era, the Noether symmetries analysis cannot be done properly. For this reason, we proposed an alternative treatment of the matter sector.
Introducing an additional homogeneous scalar field $\psi$ describing the matter, $\mathcal{L}^{\rm(m)}[g_{ab},\psi,\dot{\psi}]$, the general classification of the scalar-tensor theories is preserved, {\it i.e.} we obtain the same classification as obtained in the vacuum case~\eqref{eq:general-noether}. The same result is achieved by considering a canonical multi-field Lagrangian.

In forthcoming work, we will use the classification of viable Horndeski to implement cosmological screenings on these scalar-tensor classes characterised by the Noether symmetries and constrain the remaining free parameters.

\bigskip
\bigskip

%
%
%%%%%%%%%%%%%%%%%%%%%%%%%%%%%%%%%%%%%%%%%%%%%%

\begin{acknowledgments}
    The Authors are grateful for the support of Istituto Nazionale di Fisica Nucleare (INFN) iniziative specifiche QGSKY,  MOONLIGHT2, and TEONGRAV. M.M. thanks the University of Szczecin for the hospitality. This paper is based upon work from COST Action CA21136 {\it Addressing observational tensions in cosmology with systematics and fundamental physics} (CosmoVerse) supported by COST (European Cooperation in Science and Technology).    
\end{acknowledgments}

%%%%%%%%%%%%%%%%%%%%%%%%%%%%%%%%%%%%%%%%%%%%%
%\bibliographystyle{apsrev4-2}
\bibliography{ref}
%%%%%%%%%%%%%%%%%%%%%%%%%%%%%%%%%%%%%%%%%%%%%
\end{document}